\documentclass[a4paper,11pt]{article}
\usepackage{pos}
\usepackage{amsmath}
\usepackage{graphicx}
\usepackage{bm}

\newcommand{\be}{\begin{equation}}
\newcommand{\ee}{\end{equation}}
\newcommand{\bea}{\begin{eqnarray}}
\newcommand{\eea}{\end{eqnarray}}

\title{Multi-level computation of the hadronic vacuum polarization contribution to $(g_\mu-2)$}

\author*[a,b]{Leonardo Giusti}
\author[c]{Mattia Dalla Brida}
\author[d]{Tim Harris}
\author[b]{Michele Pepe}

\affiliation[a]{Dipartimento di Fisica, Universit\`a di Milano-Bicocca\\ Piazza della Scienza 3, 
  I-20126 Milano, Italy}

\affiliation[b]{INFN, Sezione di Milano-Bicocca \\ Piazza della Scienza 3, 
I-20126 Milano, Italy}

\affiliation[c]{Theoretical Physics Department, CERN, \\
                CH-1211 Geneva 23, Switzerland}

\affiliation[d]{School of Physics and Astronomy, University of Edinburgh, \\
                Edinburgh EH9 3JZ, UK}

\emailAdd{leonardo.giusti@mib.infn.it}
\emailAdd{mattia.dalla.brida@cern.ch}
\emailAdd{tharris@ed.ac.uk}
\emailAdd{michele.pepe@mib.infn.it}

\abstract{
The first results from the Fermilab E989 experiment have confirmed the
long-standing tension between the experimental determination of the muon
anomalous magnetic moment $a_\mu=(g_\mu-2)/2$ and its SM determination using the dispersive
approach. In order to match the expected final precision from E989, 
the current uncertainty on ab initio determinations using lattice QCD must
be decreased by a factor 5-15, a goal which is hampered by the signal-to-noise
ratio problem of the electromagnetic current correlator.
Multi-level Monte Carlo integration with fermions is a method which reduces
the variance of correlators exponentially in the distance of the fields.
Here we demonstrate that the variance reduction in a realistic two-level
simulation with a pion mass of 270 MeV, linear size of 3 fm and lattice spacing
around 0.065 fm is sufficient to compute the tail of the current
correlator with the statistical accuracy required for the hadronic vacuum
polarization contribution to $a_\mu$. An efficient estimator is also employed
for computing the disconnected contribution.

{\vspace*{3mm}CERN--TH--2021--203}
}

\FullConference{%
 The 38th International Symposium on Lattice Field Theory, LATTICE2021
  26th-30th July, 2021
  Zoom/Gather@Massachusetts Institute of Technology
}

\begin{document}
\maketitle

\section{Introduction}

The muon anomalous magnetic moment, $a_\mu$, has been measured about 15 years ago by the E821 experiment at BNL with the impressive precision of
$0.54$ parts per million (ppm)~\cite{Bennett:2006fi}. That measurement has been recently confirmed by the first data from the on-going E989
experiment at FNAL and the combination of the two results have provided the currently best experimental estimate
$a_\mu = 116 592 06.1(4.1)\times 10^{-10}$~\cite{Muong-2:2021ojo}. At the end of its operation period, the E989 experiment is expected to
attain the astonishing precision of $0.14$ ppm.

In the Standard Model (SM) the value of $a_\mu$ results from the combination of various effects that have been computed theoretically: there
are contributions from Quantum Electrodynamics and from the Weak Interactions that have been calculated perturbatively up to five loops and
two loops, respectively, as well as from the Strong Interactions coming from the Hadronic leading-order Vacuum Polarization (HVP) and the
Hadronic Light-by-Light scattering (HLbL) \cite{Aoyama:2020ynm}. The final theoretical uncertainty is currently dominated by the hadronic
part and, thus, it represents the main target to improve the accuracy of the theoretical prediction. Since the purely theoretical
computations are not sufficiently precise at the moment the hadronic contributions have been extracted (by assuming the SM) from
experimental data via dispersive integrals (HVP \& HLbL) and low-energy effective models supplemented with the operator product expansion
(HLbL). The overall theoretical expectation leads to $a_\mu = 116 591 81.0(4.3)\times 10^{-10}$ (0.37 ppm)~\cite{Aoyama:2020ynm}, which
deviates by 4.2 standard deviations from the experimental value. That difference has by now been persisting for more than a decade and it
may be a hint for New Physics.  

Although state-of-the-art lattice Quantum Chromodynamics (QCD) determinations of the HVP are steadily improving their accuracy and they are
becoming competitive, the overall error on $a_\mu$ is still 5-15 times larger than the anticipated uncertainty from E989. In fact, currently quoted
uncertainties range between $0.8\%$ to roughly $2\%$, see Ref.~\cite{Aoyama:2020ynm} and references therein. The main
difficulty in matching the level of precision of the experimental result lays in the large statistical error of
the Monte Carlo evaluation of the required correlation functions~\cite{Aoyama:2020ynm}. In a recent investigation~\cite{DallaBrida:2020cik}, we have proposed a
solution to that problem based on a multi-level Monte Carlo integration algorithm in the presence of fermions~\cite{Ce:2016idq,Ce:2016ajy}. 
The novel computational paradigm of this method with respect to the standard approach allows to reduce exponentially the variance of a
correlation function with the temporal distance of the fields. As a first application, we have performed a feasibility study on the HVP, but
the strategy is general and it can be applied to the HLbL, to the isospin-breaking and to the electromagnetic contributions as well.

\section{The problem of the signal-to-noise ratio}

A useful way of writing the HVP suitable for numerical calculations is given by

\be\label{eq:amuint}
a_\mu^{\rm HVP} = \left(\frac{\alpha}{\pi}\right)^2 \int_0^{\infty} d x_0\, K(x_0,m_\mu) \, G(x_0)\; ,
\ee

\noindent
where $\alpha$ is the electromagnetic coupling constant, $K(x_0,m_\mu)$ is a known function
increasing quadratically at large $x_0$, $m_\mu$ is the muon mass,
and $G(x_0)$ is the zero-momentum correlation function 

\be\label{eq:correlat}
 G(x_0) = \int d^3 {\bf x}\, \langle J_k^{em}(x) J_k^{em}(0) \rangle
\ee

\noindent
of two electromagnetic currents $J_k^{em}= i \sum_{i=1}^{N_f} q_i \bar\psi_i\gamma_k\psi_i$; for
unexplained notation we refer to Ref.~\cite{DallaBrida:2020cik}. In this study we consider $N_f=3$, namely the 3 lighter
quarks of QCD with the first 2 degenerate in mass, and the correlation function is given by the sum of the following contributions

\be
G(x_0) = G^{\rm conn}_{u,d}(x_0) + G^{\rm conn}_{s}(x_0) +  G^{\rm disc}_{u,d,s}(x_0)\; .  
\ee

The first and the last terms are the light-connected Wick contraction, $G^{\rm conn}_{u,d}(x_0)$, and the disconnected
one, $G^{\rm disc}_{u,d,s}(x_0)$: they represent the most problematic and numerically challenging contributions to evaluate. In fact, in
standard Monte Carlo computations, the relative error of the former at large time distances $|x_0|$ goes as  

\be\label{eq:relerr}
\frac{\sigma^2_{_{G^{\rm conn}_{\rm u,d}}}(x_0)}{[G^{{\rm conn}}_{\rm u,d}(x_0)]^2}\propto
\frac{1}{n_0}\; e^{2\, (M_\rho - M_\pi)|x_0|}\; ,  
\ee

\noindent
where $M_\rho$ is the lightest asymptotic state in the iso-triplet vector channel, and $n_0$ is the number of independent field
configurations. This shows that the exponential loss of the signal accuracy with the distance $|x_0|$ has to be compensated by a
corresponding exponential increase of the statistics $n_0$ and, thus, of the computational effort. The situation for the disconnected
contribution, $G^{\rm disc}_{u,d,s}(x_0)$, is even worse because the exponential degradation of the signal is faster since the variance is
constant in time. This exponential increase of the relative error is the bottleneck that is currently preventing to obtain a per-mille
statistical precision on the HVP by Monte Carlo simulations of QCD on the lattice.

\section{The fermionic multi-level algorithm}

\begin{figure}[!t]
\begin{center}
\includegraphics[width=12.0 cm,angle=0]{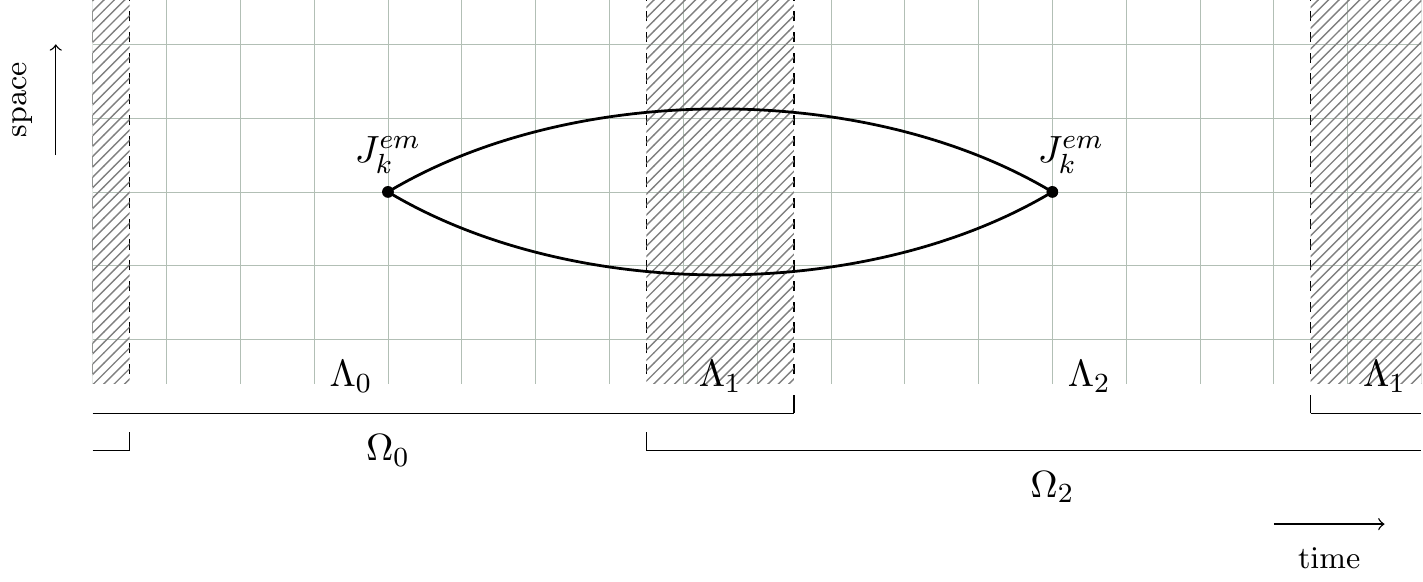}
\caption{\label{fig:MB-DD} Decomposition of the lattice in domains as described in the text;
periodic and anti-periodic boundary conditions in the time direction are enforced for gluons and fermions, respectively.}
\end{center}
\end{figure}

In the last few years an important conceptual, algorithmic and technical progress has been pursued and it is now possible to carry out
multi-level Monte Carlo simulations also in the presence of fermions~\cite{Ce:2016idq,Ce:2016ajy}. In this approach, the lattice is
initially decomposed into two overlapping domains $\Omega_0$ and $\Omega_2$ -- see e.g. Fig.~\ref{fig:MB-DD} -- which share a common region
$\Lambda_1$: it is chosen so that the minimum distance between the points belonging to the inner domains $\Lambda_0$ and $\Lambda_2$ remains
finite and positive in the continuum limit. Then, following the above decomposition, the determinant of the Hermitean massive Wilson-Dirac
operator $Q=\gamma_5 D$ is rewritten as the product of determinants of several operators

\begin{equation}\label{eq:factfinal0}
  \det\, Q= \frac{\det\, \left(1-w\right)}{\det\, Q_{\Lambda_{1}}
    \det\, Q^{-1}_{\Omega_0} \det\, Q^{-1}_{\Omega_2}} \; ,  
\end{equation}

\noindent
where $Q_{\Lambda_{1}}$, $Q_{\Omega_0}$, and $Q_{\Omega_2}$ indicate the very same operator restricted to the domains specified by the
subscript. They are obtained from $Q$ by imposing Dirichlet boundary conditions on the external boundaries of each domain. The 
matrix $w$ is built out of $Q_{\Omega_0}$, $Q_{\Omega_2}$ and the hopping terms of the operator $Q$ across the boundaries in between the inner domains
$\Lambda_0$ and $\Lambda_2$ and the common region $\Lambda_1$~\cite{Ce:2016ajy}. In the denominator the dependence on the gauge field is
already factorized because $\det Q_{\Lambda_{1}}$, $\det\, Q^{-1}_{\Omega_0}$ and $\det\, Q^{-1}_{\Omega_2}$ depend only on the gauge
field in $\Lambda_1$, $\Omega_0$ and $\Omega_2$ respectively. 

Finally, we re-express the numerator in Eq.~(\ref{eq:factfinal0}) as

\be\label{eq:MB}
\det\, \left(1-w\right) =
\frac{\det\,[1-R_{N+1}(1-w)]}{C \prod_{k=1}^{N/2} {\det} \big[ (u_k -w )^\dagger (u_k -w) \big]}\, ,
\ee

\noindent
where $u_k$ and $u^*_k$ stand for the $N$ roots of a polynomial approximant for $(1-w)^{-1}$, the numerator is the remainder, and $C$ is an
irrelevant constant. The purpose of this rewriting is that the denominator in Eq.~(\ref{eq:MB}) can be represented by an integral over a set of $N/2$ multi-boson
fields~\cite{DallaBrida:2020cik,Ce:2016idq,Ce:2016ajy} having an action with a factorized dependence on the gauge field in $\Lambda_0$ and
$\Lambda_2$ inherited from $w$. By a proper choice of the polynomial approximation, the remainder in the numerator of Eq.~(\ref{eq:MB})
fluctuates mildly with the gauge field and it can be included in the observable as a reweighting factor.

In Fig.~\ref{fig:MB-DD} we show a convenient decomposition of the lattice: the two regions $\Lambda_0$ and $\Lambda_2$ have the shape of thick
time-slices while $\Lambda_1$ includes the rest of the lattice and it has the shape of thinner time-slices that keep the first two regions apart. 
Taking into account the short-distance suppression of the quark propagator, a thickness of $0.5$~fm or so for the slices forming $\Lambda_1$
is good enough to make $\Lambda_0$ and $\Lambda_2$ weakly correlated, furthermore it is not expected to vary significantly with the quark
mass. This is the domain decomposition that we use for the numerical computations presented here.

A two-level scheme is used in the Monte Carlo simulation. As a first step, $n_0$ well-decorrelated, level-$0$ gauge field configurations are
generated by updating the field over the whole lattice: they represent the starting point of the update at level-$1$. This second step is
accomplished as follows: in every level-$0$ configuration the gauge field in the overlapping region $\Lambda_1$ is kept fixed and $n_1$
level-$1$ configurations are generated by updating the field in $\Lambda_0$ and in $\Lambda_2$ independently thanks to the factorization of the action.  
The resulting gauge fields are then combined in all possible ways obtaining effectively $n_0\cdot n_1^2$ configurations at the cost of generating $n_0\cdot n_1$
gauge fields over the entire lattice. Past experience on two-level integration suggests that, with two independently updated regions,
the variance decreases proportionally to $1/n_1^2$ until the standard deviation of the estimator is comparable with the signal, i.e. until
the level-$1$ integration has solved the signal-to-noise problem. From Eq.~(\ref{eq:relerr}) we thus conclude that, as expected, the variance reduction due
to level-$1$ integration grows exponentially with the time-distance of the currents in Eq.~(\ref{eq:correlat}).

\section{The numerical study}

In this section we present the results of a study on the efficiency of the two-level Monte Carlo algorithm. We have performed numerical
simulations of QCD with two dynamical flavours supplemented by a valence strange quark on a lattice of size $96\times 48^3$
with a spacing of $\;a=0.065$\,fm, and with a pion mass of $270$~MeV. The lattice has been decomposed such that the domains $\Lambda_0$ and
$\Lambda_2$ are made of $40$ consecutive time-slices separated by two regions of $8$ time-slices each which form the region $\Lambda_1$.
We use the standard pseudofermion representation for the determinants at the denominator of Eq.~(\ref{eq:factfinal0}) while the number of
multi-bosons is fixed to $N=12$. We consider the same action and the same set of auxiliary fields both at level-$0$ and at level-$1$.
The reweighting factor is estimated stochastically with 2 random sources which are sufficient for making its contribution
to the statistical error negligible. At level-0 we have generated $n_0=25$ configurations and then, for each of them, $n_1=10$
configurations in $\Lambda_0$ and in $\Lambda_2$ have been produced. In Refs.~\cite{DallaBrida:2020cik,Ce:2016idq,Ce:2016ajy} one can find further
details on the algorithm and its implementation.

\begin{figure*}[htb]
\begin{center}
\begin{tabular}{cc}
\includegraphics[width=7.5 cm,angle=0]{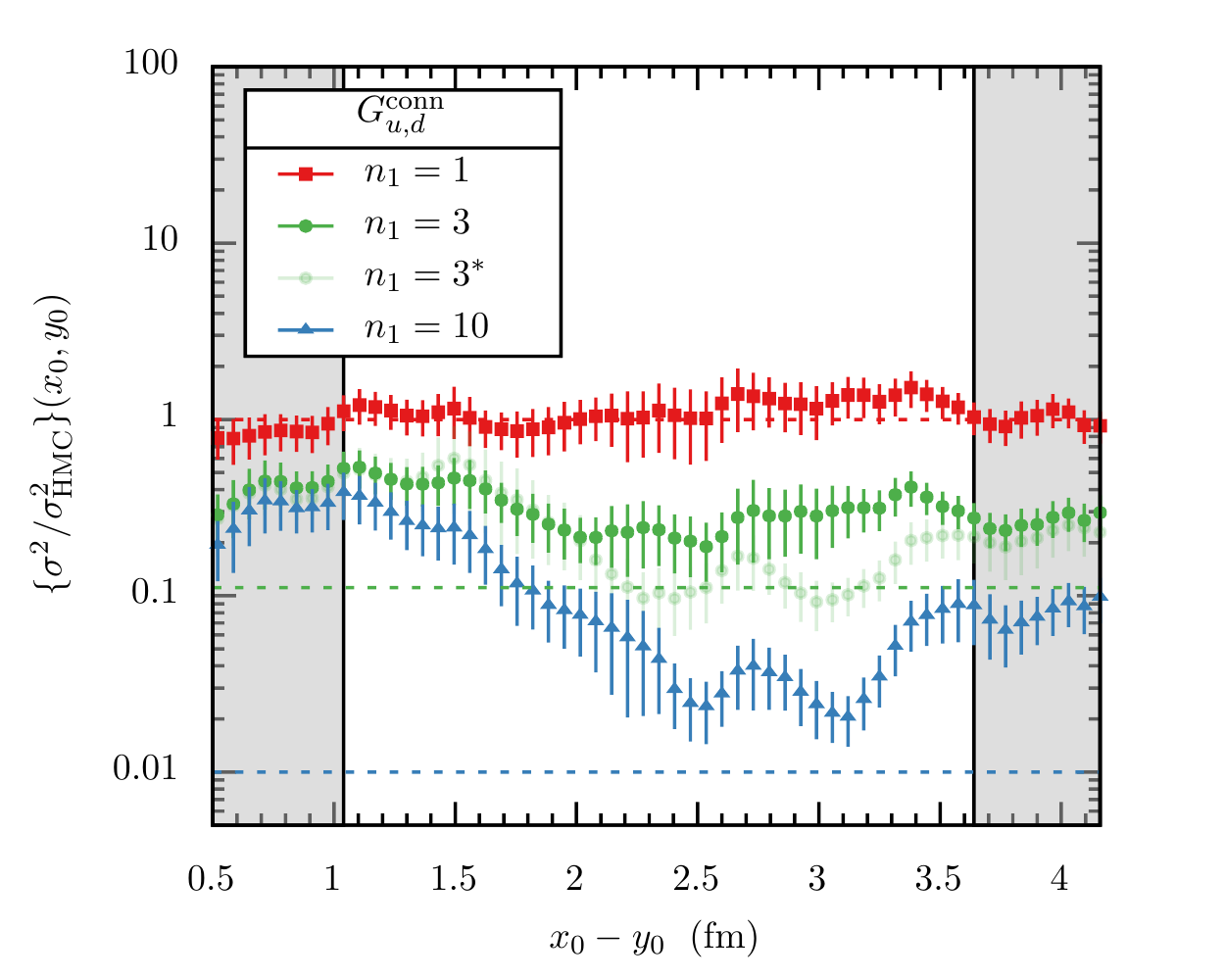} &
\includegraphics[width=7.5 cm,angle=0]{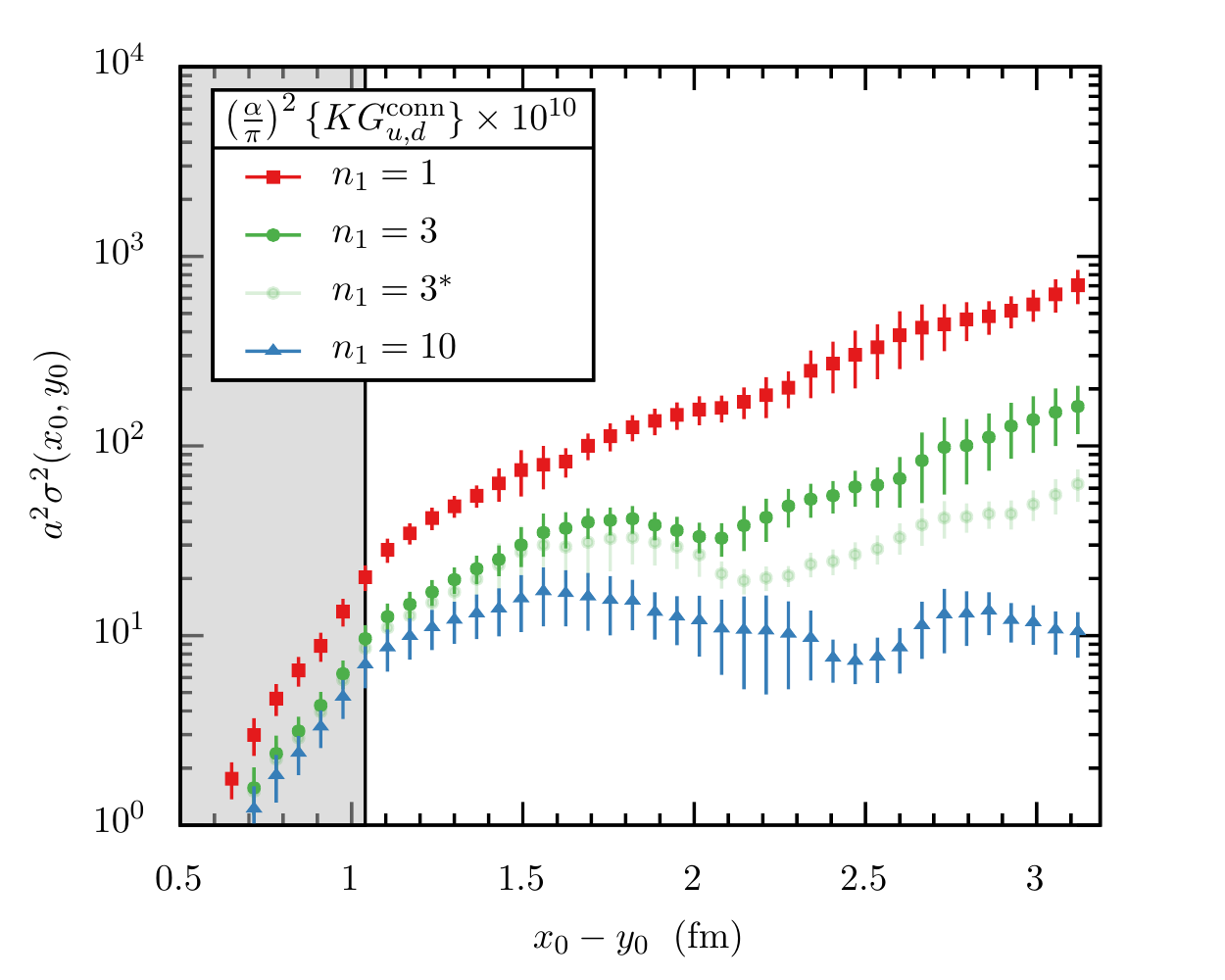}\\
\end{tabular}

\caption{Left panel: dependence of the variance of the light-connected contraction on the difference between the time-coordinates of the
  currents for $n_1=1,3,10$. Data are normalized to the analogous ones computed on CLS
configurations generated by one-level HMC. Dashed lines represent the maximum reduction, $1/n^2_1$, that can be obtained by the
two-level integration with decorrelated level-$1$ configurations. Grey bands indicate the thick time-slices
where the gauge field is kept fixed during level-$1$ updates. Right panel: variance of the light-connected
contribution to the integrand in Eq.~(\ref{eq:amuint}).
\label{fig:variances}}
\end{center}
\end{figure*}

We have carried out a dedicated calculation of the correlation functions in order to assess the reduction of the variance due only to
two-level averaging. The light-connected contraction has been calculated by averaging over $216$ local sources located on the time-slice $y_0/a=32$ of
$\Lambda_0$ -- corresponding to a distance of $8$ lattice spacings from its right boundary -- and by summing over the sink space-position. 
For what concerns the disconnected contraction, each single-propagator trace has been averaged over 768 Gaussian random sources so to have a
negligible random-noise contribution to the variance~\cite{DallaBrida:2020cik,Giusti:2019kff}.

In the left panel of Fig.~\ref{fig:variances} we show the variance of the light-connected contribution as a function of the distance from the source:
for sake of readability only the time-slices belonging to $\Omega_2$ are displayed, i.e. those relevant for studying the effect of two-level integration
given the source position. Data are normalized to the variance obtained with the same number of sources on CLS configurations which were
generated with a conventional one-level HMC. The data clearly provide evidence of the exponential reduction of the variance with the
distance from the source with the maximum gain reached from $2.5$~fm onward for $n_1=10$. We observe a mild reduction with respect to the
ideal scaling $n_1^2$ by a factor between $2$ and $3$ either for $n_1=3$ or $10$ (dashed lines): this could be related to a residual
correlation among level-$1$ configurations. The power of the two-level integration is also shown in the right panel of
Fig.~\ref{fig:variances}, where the variance of the light-connected contribution to the integrand in Eq.~(\ref{eq:amuint}) is plotted
as a function of the time-distance of the currents. The sharp rising of the variance computed by one-level Monte Carlo ($n_1=1$, red squares)
is automatically flattened out by the two-level multi-boson domain-decomposed HMC ($n_1=10$, blue triangles)
without the need for modeling the long-distance behaviour of $G^{\rm conn}_{u,d}(x_0)$.

Finally, we compute the dependence of the integral in Eq.~(\ref{eq:amuint}) on the upper extrema of integration $x_0^{\rm max}$: for $n_1=1$, the
integral reads $446(26)$ and $424(38)$ for $x_0^{\rm max}=2.5$ and $3.0$~fm respectively, while for $n_1=10$ the corresponding values are
$467.0(8.4)$ and $473.4(8.6)$. It is interesting to note that with the one-level integration the errors on the contributions to the integral
from $0$ to $2.5$~fm and from $2.5$ to the maximum value of $3.0$~fm are comparable, while with the two-level HMC the contribution to the
variance from the long distance part becomes negligible. Similar remarks hold for the much smaller disconnected contribution.

\section{Results and discussion}

\begin{figure*}[!t]
\begin{center}
\begin{tabular}{cc}
\includegraphics[width=7.5 cm,angle=0]{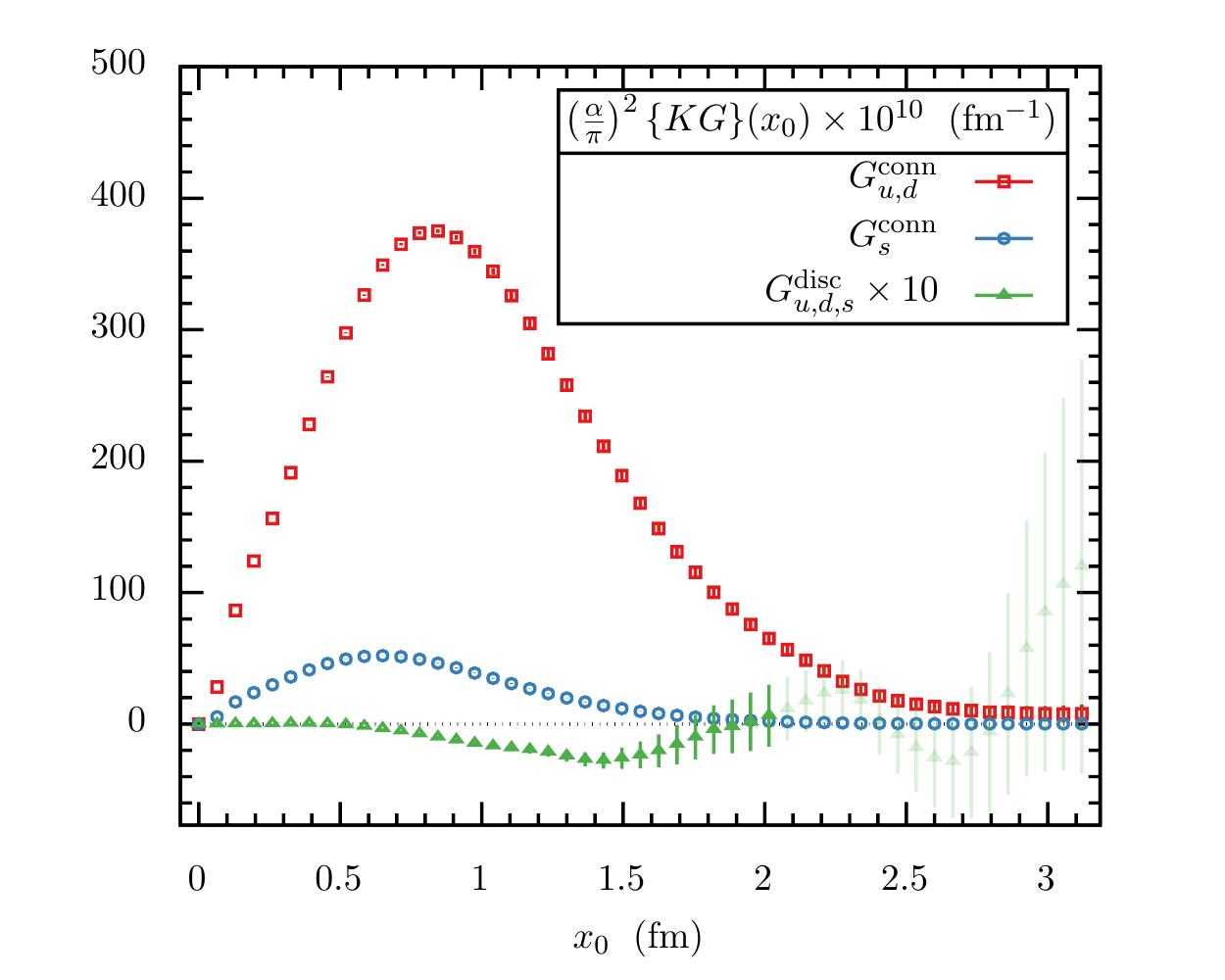} &
\includegraphics[width=7.5 cm,angle=0]{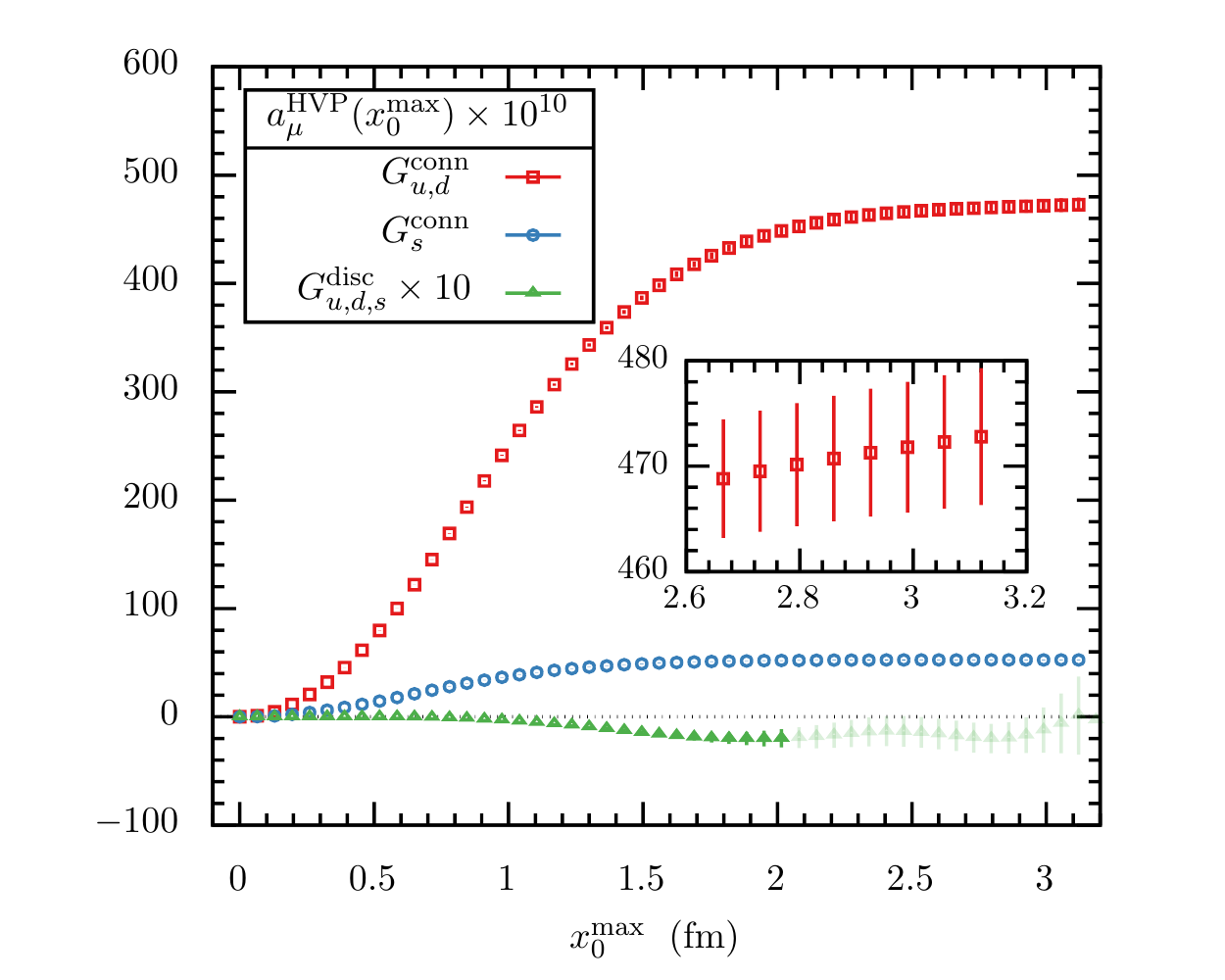}\\
\end{tabular}

\caption{Best results for contribution of the light-connected (red squares), strange-connected (blue circles) and disconnected (green triangles)
  contractions to the integrand in Eq.~(\ref{eq:amuint}) (left panel) and to $a_\mu^{\rm HVP}$ (right panel) as a function of the time
  coordinate $x_0^{\rm max}$.
\label{fig:integrand}}
\end{center}
\end{figure*}

On the l.h.s. of Fig.~\ref{fig:integrand} our best result for the light-connected contribution to the integrand in Eq.~(\ref{eq:amuint}) are
plotted with red squares. These results are obtained by a weighted average of the above discussed correlation function computed 
on $32$ point sources per time-slice on $7$ time-slices at $y_0/a=\{8,16,24,56,64,72,80\}$ and on $216$ sources at $y_0/a=32$.
The statistical accuracy is good up to the maximum distance of $3$~fm or so. The contribution coming from the connected contraction of the
strange quark, $G^{\rm conn}_{s}(x_0)$, is much less noisy, and it is accurately determined by averaging over $16$ point sources at $y_0/a=32$. 
It is at most one order of magnitude smaller than the light-connected one and the impact of the statistical error on the final accuracy is negligible. 
The data are shown using blue circles in the left plot of Fig.~\ref{fig:integrand}. The disconnected contribution has been computed as
discussed in the previous section and the numerical results are shown in the left plot of Fig.~\ref{fig:integrand} using green triangles.
We observe a negative peak at about $1.5$~fm, and a good statistical signal is obtained up to $2.0$~fm or so. Its absolute value is more
than two orders of magnitude smaller than the light-connected contribution over the entire range explored.

The best values of the light-connected (red squares), of the strange-connected (blue circles), and of the disconnected (green triangles)
contributions to $a_\mu^{\rm HVP}\cdot 10^{10}$ are shown in the right plot of Fig.~\ref{fig:integrand} as a function of the upper extrema
of integration $x_0^{\rm max}$ in Eq.~(\ref{eq:amuint}). The light-connected part starts to flatten out at $x_0^{\rm max} \sim 2.5$~fm and,
at the conservative distance of $x_0^{\rm max}= 3.0$~fm, its value is $471.8(6.2)$. At the same distance the value of the strange-connected
contribution is $52.55(21)$ with a negligible error with respect to the light-connected one. For what concerns the disconnected
contribution, it starts to flatten out at about $x_0^{\rm max} \sim 2.0$~fm, where its value is $-1.98(84)$. For $x_0^{\rm max} =3.0$~fm,
its statistical uncertainty is $2.1$ which is still 3 times smaller with respect to the light-connected one. Although the disconnected
contribution is very small it must be taken into account to attain the target per-mille precision on the HVP; the combined usage of split-even
estimators \cite{Giusti:2019kff} and two-level integration solves the problem of its computation that has represented a numerical challenge
for quite some time. By combining the connected contributions at $x_0^{\rm max}=3.0$~fm with the disconnected part at
$x_0^{\rm max}=2.0$~fm, the best final estimate is $a_\mu^{\rm HVP} = 522.4(6.2) \cdot 10^{-10}$.  

In this investigation we demonstrate the effectiveness of the fermionic multi-level algorithm by achieving a 1\% statistical precision with just
$n_0\cdot n_1=250$ configurations on a realistic lattice. This shows that, for the light-quark mass considered here, a per-mille statistical precision
on $a_\mu^{\rm HVP}$ is reachable by increasing $n_0$ and $n_1$ by a factor of about $4$--$6$ and $2$--$4$ respectively. For lighter up and
down quarks, the gain due to the multi-level integration is expected to increase exponentially in the quark mass, hence improving even more
dramatically the scaling of the simulation cost with respect to a standard one-level Monte Carlo. In conclusion, the change of computational
paradigm discussed here overcomes the main barrier that prevents to reach a per-mille precision on $a_\mu^{\rm HVP}$ on computers available today.

\section*{Acknowledgments}  
The generation of the configurations and the measurement of the correlators have been performed on the
PC clusters Marconi at CINECA (CINECA- INFN, CINECA-Bicocca agreements, ISCRA B project HP10BF2OQT)
and at the Juelich Supercomputing
Centre, Germany (PRACE project n. 2019215140) while the R\&D has been carried out on Wilson and Knuth at
Milano-Bicocca. We thank these institutions and PRACE for the computer resources and the
technical support. We also acknowledge PRACE for awarding us access to MareNostrum at Barcelona Supercomputing
Center (BSC), Spain (n. 2018194651) where comparative performance tests of the code have been performed.
We acknowledge partial support by the INFN project ``High performance data network''.

\end{document}